\documentclass[11pt,a4paper]{article}

\usepackage{fullpage}
\usepackage{indentfirst}
\usepackage{color}
\usepackage[utf8x]{inputenc}
\usepackage[T1]{fontenc}
\usepackage[active]{srcltx}
\usepackage{hyperref}
\usepackage[english]{babel}
\usepackage{amsmath}
\usepackage{amsfonts}
\usepackage{amssymb}
\usepackage{amscd}
\def \be{\begin{equation}}
\def \ee{\end{equation}}
\def \nn{\nonumber}
\def \noi{\noindent}

\def \d{{\rm d}}
\newcommand{\parcial}[1]{ \frac{\partial}{\partial #1} }
\newcommand{\dparcial}[2]{ \frac{\partial #1}{\partial #2} }

\author{V. Aldaya$^{1}$, F. Coss\'{\i}o$^{1}$, J. Guerrero$^{1,2}$ and F. F.
L\'opez-Ruiz$^{1}$}

\title{A symmetry trip from Caldirola to Bateman damped systems}

\date{\begin{center}
\begin{small}$^1$ Instituto de Astrof\'{\i}sica de Andaluc\'{\i}a, IAA-CSIC, 
\end{small}\\	
\begin{small}
  Apartado Postal 3004, 18080 Granada, Spain
\end{small}\\
\begin{small}$^2$Departamento de Matem\'atica Aplicada, Universidad de
Murcia, \end{small}\\
\begin{small}Campus de Espinardo, 30100 Murcia, Spain.\end{small}\\
\begin{small}valdaya@iaa.es\ fcossiop@gmail.es \ juguerre@um.es\ flopez@iaa.es
\end{small}\\                                                                   
\end{center}
}
\begin{document}

\maketitle
%

%\tableofcontents

\begin{abstract}
% 
% In a previous paper it was shown that a basic, Heisenberg-Weyl algebra of
% conserved quantum operators can be imported from the free particle system to
% the
% damped harmonic oscillator described by the Caldirola-Kanai model by means of
% a
% quantum version of the Arnold transformation of classical mechanics.
% 
For the Caldirola-Kanai system, describing a quantum damped harmonic
oscillator, a couple of constant-of-motion operators generating the Heisenberg
algebra can be found.
The inclusion of the standard time evolution symmetry in this algebra for damped
systems, in a unitary manner, requires a non-trivial extension of this basic
algebra and hence the physical system itself. Surprisingly, this
extension leads directly to the so-called Bateman's dual system, which now 
includes a new particle acting as an energy reservoir.
The group of symmetries of the dual system is presented, as well as a
quantization that implies, in particular, a first-order Schr\"odinger equation.
The usual second-order equation and the inclusion of the original
Caldirola-Kanai model in Bateman's system are also discussed. 

\end{abstract}
%\tableofcontents{}

\section{Introduction}

%Poner relación con los operadores de creación y aniquilación de
%Feschbach-Ticochinsky (recordar que la idea de definir esos a y acruz es
%resolver el problema de autovalores).

The interest in dissipative systems at the quantum level has remained constant
since the early days of Quantum Mechanics. The difficulties in describing
damping, which intuitively could be understood as a mesoscopic property, within
the fundamental quantum framework, have motivated a huge amount of papers.

Applications of quantum dissipation abound. For example, in quantum optics,
where the quantum theory of lasers and masers makes use of models including
damping \cite{Scully-Zubairy}, or in the study of decoherence phenomena
\cite{Breuer}. Some authors have modeled dissipation by means of the theory of
open systems or the thermal bath approach, in which a damped system is
considered to be a subsystem of a bigger one  with infinite degrees of freedom
\cite{Dittrich,Breuer}. However, damped systems are interesting in themselves as
fundamental ones. In particular, the quantum damped harmonic oscillator,
frequently described by the Caldirola-Kanai equation \cite{Caldirola,Kanai}, has
attracted much attention, as it could be considered one of the simplest and
paradigmatic examples of dissipative system.

The description of the quantum damped harmonic oscillator by the Caldirola-Kanai
model, which includes a time-dependent Hamiltonian, has been considered to have
some flaws. For instance, it is claimed that uncertainty relations are not
preserved under time evolution and could eventually be violated
\cite{Brittin,Razavy_libro}. Many considerations were made in this direction.
For example, Dekker in \cite{Dekker_articulo} introduced complex variables and
noise operators to tackle the problem, claiming that no dynamical description in
terms of a Schr\"odinger wave function can be expected to exist. In
\cite{Razavy}, a non-linear Schr\"odinger-Langevin wave equation was proposed as
the starting point in formulating the quantum theory. However, this
inconsistency seems to be associated  with a confusion between canonical
momentum and ``physical'' momentum  \cite{Schuch}. 

Despite these considerations about the Caldirola-Kanai model, many developments
went ahead. Coherent states were calculated in \cite{Manko} by finding creation
and annihilation operators, built out of operators which commute with the
Schr\"odinger equation. The corresponding number operator turns out to be an
auxiliary, conserved  operator, obviously different from the time-dependent
Hamiltonian. This paper also defined the so-called loss energy states for the
damped harmonic oscillator. The famous report by Dekker \cite{Dekker_report}
provides a historical overview of some relevant results.

The analysis of damping from the symmetry point of view has proved to be
especially fruitful. In a purely classical context, the symmetries of the
equation of the damped harmonic oscillator with time-dependent parameters were
found in  \cite{Martini}. Two comprehensive articles,
\cite{Cervero-C,Cervero-Q}, are of special interest. In those papers the authors
found, for the damped harmonic oscillator, finite-dimensional point symmetry
groups for the corresponding Lagrangian (the un-extended Schr\"odinger group
\cite{schrogroup}) and the equations of motion ($SL(3,\mathbb{R})$)
respectively, and an infinite contact one for the set of trajectories of the
classical equation. They singled out a ``non-conventional'' Hamiltonian from
those generators of the symmetry, recovering some results from \cite{Manko}.
Then, they concluded that the damped harmonic oscillator should not be claimed
to be dissipative at all at the quantum level, since this 
``non-conventional'' Hamiltonian is conserved and even related to an
oscillator with variable frequency. In any case, it  still remains to address the 
symmetry role of the time translation generator $i\hbar \parcial{t}$. In fact, 
$i\hbar \parcial{t}$ acting on a solution of the Schr\"odinger equation is no longer
a solution. As a consequence, the time evolution 
operator $\hat{U}$ does not constitute a uniparametric group of unitary transformations.
 Equivalently, the solution of the equation 
$i\hbar\parcial{t}\hat{U}=\hat{H}(t)\hat{U}$ is not $e^{\frac{i}{\hbar}t \hat{H}(t)}$, but
rather the time ordered product $Te^{-\frac{i}{\hbar}\int\hat{H}(t)dt}$, referring to the Neuman series,  or the Magnus
series $e^{-\frac{i}{\hbar}\hat{\Omega}(t)}$ \cite{Magnus}.

Many papers related to the Caldirola-Kanai model keep appearing, showing that
the debate about fundamental quantum damping is far from being closed.
We can mention \cite{Yeon1987}, where the driven damped harmonic
oscillator is analyzed, or the review \cite{Um-Yeon}. Even the possible
choices of classical Poisson structures and Hamiltonians, or generalizations to
the non-commutative plane, have deserved attention as recently as in
\cite{Hone} and \cite{Streklas}, respectively.

In fact, in \cite{arnoldnuestro}, the authors provided a neat framework to study
this model, based on a quantum generalization of the Arnold transformation
\cite{Arnold}. The integrals of motion and symmetries were identified and
exploited to calculate wave functions, basic operators and the exact time
evolution operator.

Besides the Caldirola-Kanai model, the Bateman's dual system appears as an
alternative description of dissipation in the damped harmonic oscillator. 
In his original paper \cite{Bateman}, Bateman looked for a variational
principle for equations of motion with a friction term linear in velocity, but
he allowed the presence of extra equations. This trick effectively doubles the
number of degrees of freedom, introducing a time-reversed version of the
original damped harmonic oscillator, which acts as an energy reservoir and could
be considered as an effective description of a thermal bath. The Hamiltonian
that describes this system was rediscovered by Feschbach and Tikochinsky
\cite{Tikochinsky,Fesh-Tiko,Feschbach_libro,Dekker_report} and the corresponding
quantum
theory was immediately analyzed.

Some issues regarding the Bateman's system arose. The Hamiltonian presents
a set of complex eigenvalues of the energy (see \cite{Chruscinski} and
references therein), and the vacuum of the theory decays with time. This last
feature was treated in \cite{Celeghini}, where Celeghini et al. suggested that
the quantum theory of the dual system could find a more natural framework in
quantum field theory\footnote{We feel that the ultimate reason is nevertheless
the lack of a vacuum representation of~the~relevant~group~(see~Section~\ref{sec:arnold.newgroup}).}. 
On the other hand, in \cite{Chruscinski} the
generalized
eigenvectors corresponding to the complex eigenvalues are interpreted as
resonant states.

Bateman's dual system is still frequently discussed \cite{Blasone}. Many authors
have considered this model as a \textit{good} starting point for
the formulation of the quantum theory of dissipation. One of the aims of this
chapter will be to show that the study of the symmetries of the Caldirola-Kanai
model leads to the Bateman's dual system, thus to be considered as a
\textit{natural} starting point for the study of quantum dissipation.

The purpose of this article is to throw some light on the subject of quantum
dissipation by putting together all those bricks with the guide of symmetry.
We begin in Section~\ref{introarnold} by recalling the results from
\cite{arnoldnuestro}, in the case of the damped harmonic oscillator with
constant coefficients. In particular, by using the quantum Arnold
transformation, we import basic operators from the free particle system, which
satisfy the condition of being integrals of the motion and close a
Heisenberg algebra. Also, the complete set of symmetries of the quantum free
particle, the Schr\"odinger group, can be realized on the Caldirola-Kanai model,
providing as many conserved quantities as in the free particle.

Time translations in the non-free system do not belong the imported
Schr\"odinger group. This is to be expected, as the classical equation of motion
includes a friction term and the energy in this system is not conserved. The
following question immediately arises: Is there any finite-dimensional group of
symmetry containing time translations and, at least, the basic operators? The
answer is `yes', and Section~\ref{sec:arnold.newgroup} pays attention to this question in
the case of the damped harmonic oscillator and the surprising consequences of
the subsequent calculation: for this symmetry to act properly, it is necessary
to enlarge the physical system with a new degree of freedom, corresponding to a
new particle with interesting properties. This could be understood as a very
simple version of the gauge principle, in which a bigger symmetry for the
original ``free'' system is imposed. In fact, this new system with two
degrees of freedom \textit{is} the Bateman's dual system.
Taking advantage of the symmetry approach, we will go a bit further and provide
a group law corresponding to the symmetries of the dual system
(Section~\ref{bategroup}). 

With the light of this group law, in Section~\ref{sec:arnold.quantbat} we give an
analysis of the quantization of the dual system that we have encountered. In
particular, we show that it is possible to find a first-order Schr\"odinger
equation (Subsection~\ref{firstorderpola}), from which the wave functions and
the energy spectrum can be obtained, as well as the more usual second-order
equation. In Subsection~\ref{back2caldi} we illustrate how the Caldirola-Kanai
system can be recovered by means of a constraint process.
An Appendix is devoted to the study of an infinite dimensional symmetry algebra for the damped particle.

\section{Basic operators in the Caldirola-Kanai model}
\label{introarnold}

Let us first introduce the Caldirola-Kanai equation, which is a Schr\"odinger
equation for the Damped Harmonic Oscillator (DHO):              
\begin{equation}
i\hbar\frac{\partial \phi}{\partial t} = \hat H_{DHO}\,\phi \equiv
-\frac{\hbar^{2}}{2m} e^{-\gamma t}\frac{\partial^{2} \phi}{\partial x^{2}}
+\frac{1}{2}m\omega^{2}x^{2} e^{\gamma t}\phi \,,
\label{eq:Schrodinger DHO}
\end{equation}
where $\gamma$ and $\omega$ are constants defining the system. It is derived by
standard canonical quantization from a time-dependent Hamiltonian whose
quantum operator is $\hat H_{DHO}$. The corresponding classical equation
of motion is
\begin{equation}
\ddot{x}+\gamma\dot{x}+\omega^{2}x=0\,.
\label{eq:ecuacionDHO}
\end{equation}

There are different ways to identify basic position and momentum operators
associated with classical conserved quantities (Noether invariants). In general a conserved quantum operator $\hat O(t)$
must satisfy the relation:
\begin{equation}
  \frac{\d}{\d t} \hat O(t) \equiv \frac{\partial}{\partial t} \hat O(t) 
   + \frac{i}{\hbar}[\hat H(t),\hat O(t)] = 0\,.
\end{equation}

We find particularly interesting the Quantum Arnold Transformation (QAT) technique developed in \cite{arnoldnuestro}. 

The QAT (or, rather, the
inverse) relates the Hilbert space $\mathcal H^G_\tau$ of
solutions of the  Schr\"odinger equation  for the Galilean particle
\begin{equation}
i\hbar\frac{\partial \varphi}{\partial \tau}=-\frac{\hbar^{2}}{2m}
\frac{\partial^{2} \varphi}{\partial \kappa^{2}}\, ,
\label{eq:Schrodinger libre}
\end{equation}
to the corresponding Hilbert space $\mathcal H_t$ of the DHO. The QAT can be written as
\begin{equation}
\begin{split}
  \hat A:& \quad \mathcal H_t \;\; \longrightarrow \quad \mathcal H^G_\tau \\
    & \phi(x,t)  \longmapsto \; \varphi(\kappa,\tau) = 
        \hat A \bigl( \phi(x,t) \bigr) = 
       A^* \bigl( \sqrt{u_{2}(t)}\,e^{-\frac{i}{2}\frac{m}{\hbar}
                 \frac{1}{W(t)}\frac{\dot{u}_{2}(t)}{u_{2}(t)}
                                                 {x}^{2}} \phi(x,t) \bigr) \,,
\end{split}
\end{equation}
where $A^*$ denotes the pullback operation corresponding to the classical Arnold transformation $A$ \cite{Arnold}:
\begin{equation}
\begin{split}
  A:& \; \; \mathbb R \times T \longrightarrow \mathbb R \times T' \\
    & \quad (x,t)  \longmapsto \; (\kappa,\tau) =
        A\bigl((x,t)\bigr) = (\tfrac{x}{u_2},\tfrac{u_1}{u_2})\,,
\end{split}
\end{equation}
and $T$ and $T'$ are open intervals of the real line containing $t=0$
and $\tau=0$, respectively. Here $u_1(t)$ and $u_2(t)$ are
 independent solutions of (\ref{eq:ecuacionDHO}), satisfying
the initial conditions  $u_1(0)=0,  u_2(0)=1, \dot{u}_1(0)=1, \dot{u}_2(0)=0$ and  
$W(t) \equiv \dot{u}_{1}(t) u_{2}(t) - u_{1}(t)\dot{u}_{2}(t)$ (see \cite{arnoldnuestro}). For the DHO they are:
\begin{equation}
u_{1}(t)=\frac{1}{\Omega}e^{-\frac{\gamma}{2}t}\sin\Omega t,\qquad
u_{2}(t)=e^{-\frac{\gamma}{2}t}\cos\Omega t + 
\frac{\gamma}{2 \Omega}e^{-\frac{\gamma}{2}t}\sin\Omega t,
\label{eq:us DHO}
\end{equation}
for which  $W(t)= e^{-\gamma t}$, and                                      
\begin{equation}
\Omega=\sqrt{\omega^{2}-\frac{\gamma^{2}}{4}}\,.
\label{eq:frecuencia}
\end{equation}
Note that these solutions have good limit in the case of critical
damping $\omega = \tfrac{\gamma}{2}$.

%Then, we go from

As already remarked, the basic symmetries of the free system are inherited by
the DHO system, and we are now able to transform the infinitesimal
generators of translations (the Galilean momentum operator $\hat \pi$,
corresponding to the classical conserved quantity `momentum') and
non-relativistic boosts (the position operator $\hat \kappa$, corresponding to
the classical conserved quantity `initial position'). They are, explicitly,
\begin{align}
\hat \pi &= -i \hbar \frac{\partial}{\partial \kappa} 
\label{eq:operata momento libre}
\\
\hat \kappa &= \kappa + \frac{i \hbar}{m} \tau
\frac{\partial}{\partial \kappa}\;, 
\label{eq:operata posicion libre}
\end{align}
that is, those basic, canonically commuting operators with constant expectation
values, that respect the solutions of the free Schr\"odinger
equation, have constant matrix elements  and fall down to well defined, time-independent operators in the
Hilbert space of the free particle $L^2(\mathbb R)$,  $\mathcal H^G_{\tau=0}$.

The basic quantum operators, as derived by means of the inverse QAT on $\hat \pi$ and $\hat \kappa$ are:
\begin{align}
 \hat P &= 
 -i \hbar \frac{e^{-\frac{\gamma t}{2}}}{2 \Omega }(2 \Omega \, \cos \Omega t
   +\gamma \, \sin \Omega t) \frac{\partial}{\partial x} 
   + m \frac{e^{\frac{\gamma t}{2}} }{4 \Omega} 
      \left(\gamma ^2+4 \Omega^2\right) \,\sin \Omega t \, x\,,
\label{eq:P_DHO}
 \\
 \hat X &= 
  \frac{e^{\frac{\gamma t}{2}}}{2 \Omega } (2 \Omega \, \cos \Omega t
   -\gamma \, \sin \Omega t) \, x
    +i \hbar \frac{e^{-\frac{\gamma t}{2}}}{m \Omega }\, \sin \Omega t
     \frac{\partial}{\partial x}\,.
\label{eq:X_DHO}
\end{align}

with

\begin{equation}
  \left[ \hat X,\hat P \right] = i \hbar \,.
\end{equation}

% Even though it is possible to set up a clear framework to deal with any
% LSODE-type quantum system by employing the quantum Arnold transformation, it
% does not provide by itself a well-defined operator associated with proper time
% evolution. This is rooted in the fact that this conventional time evolution is
% not included in the symmetry group that can be imported from the free system:
% the Hamiltonian does not belong to the specific representation of the
% Schr\"odinger algebra. One may wonder what happens if time evolution symmetry
% is forced. We shall pursue this issue for the damped harmonic oscillator in the
% next Section.

\section{Deriving dissipative forces from a symmetry}
\label{sec:arnold.newgroup}

Even though it is possible to set up a clear framework to deal with the quantum DHO
 system by employing the QAT, it
does not provide by itself a well-defined operator associated with the actual time
evolution. As mentioned in the Introduction, this is rooted in the fact that the conventional time evolution is
not included in the symmetry group that can be imported from the free system:
the Hamiltonian does not preserve the Hilbert space of solutions of the DHO Schr\"odinger equation. 
One may wonder what happens if time evolution symmetry
is forced. We shall pursue this issue for the damped harmonic oscillator in this 
Section.

%%%%%%%%%%%%%%%%%%%%%%%%%%%%%%%%%%%%%%%%%%%%%%%%%%%%%%%%%%%%%%%%%%%%%%%%%%%%%%%%%%%%%%%%%%%%%%%%%%%%%%%%%%%%%%%%%%%%%%%%%5
\subsection{Time symmetry}
\label{timesymmetry}

Historically, Caldirola and Kanai derived their Hamiltonian from the Bateman one
by means of time-dependent canonical transformations. 
Now 
we are going to proceed in the opposite
direction, deriving Bateman Hamiltonian from Caldirola-Kanai one by purely
symmetry considerations.

In the damped harmonic oscillator, neither the operator
$i\hbar\frac{\partial}{\partial t}$, nor $\hat H_{DHO}$ (which coincides with
the former on solutions) close under commutation with  $\hat X$ and $\hat P$
(see equations \eqref{eq:X_DHO} and \eqref{eq:P_DHO}). 
We will impose the condition for the time translation to be a symmetry. But
will do it in an elegant way, trying to close an algebra of (constant, symmetry
generating) observables, taking advantage of the expressions
for the basic operators obtained by the quantum Arnold transformation. 
So, we wonder if it is possible to incorporate
$i\hbar\frac{\partial}{\partial t}$
into the basic Lie algebra of operators, trying to close an enlarged Lie
algebra acting on
the (possibly enlarged) Hilbert space ${\cal H}_t$.
The answer to this question is in the affirmative, but it requires a delicate 
analysis. The resulting enlarged algebra will include $\hat{X},
\hat{P}, \hat{H} \equiv  i\hbar\frac{\partial}{\partial t}$ and four more
generators (plus the central one $\hat{I}$), denoted by $\hat{Q}, \hat{\Pi},
\hat{G}_1$ and $\hat{G}_2$\footnote{In the simpler case of the damped
particle, infinitely many new generators can be included in its Lie algebra.
See Appendix~\ref{infinitegroup} for further details.}. 

Together with the generators $\hat X$ and $\hat P$ and the Hamiltonian, let us
introduce the following operators: 
\begin{align*}
 \hat P &= 
 -i \hbar e^{-\frac{\gamma t}{2}} (\cos \Omega t
   +\frac{\gamma}{2 \Omega} \, \sin \Omega t) \frac{\partial}{\partial x} 
   + m \,\frac{\omega^2}{\Omega} \,e^{\frac{\gamma t}{2}} 
       \,\sin \Omega t \, x\,,
 \\
 \hat X &= 
  e^{\frac{\gamma t}{2}} (\cos \Omega t
   -\frac{\gamma}{2 \Omega} \, \sin \Omega t) \, x
    +i \hbar \frac{e^{-\frac{\gamma t}{2}}}{m \Omega }\, \sin \Omega t
     \frac{\partial}{\partial x}\,
 \\
 \hat \Pi &= 
    i \hbar e^{-\frac{\gamma t}{2}}\, 
    (\cos \Omega t - \frac{\gamma}{2\Omega}\, \sin \Omega t)
    \frac{\partial}{\partial x}
   - m \frac{\omega^2}{\Omega } \,e^{\frac{\gamma t}{2}}\, 
    \sin \Omega t \, x
 \\
 \hat{\tilde Q} &=
  e^{\frac{\gamma t}{2}} (\cos \Omega t
   -\frac{3 \gamma}{2 \Omega} \, \sin \Omega t) \, x
    +i \hbar \frac{e^{-\frac{\gamma t}{2}}}{m \Omega }\, \sin \Omega t
     \frac{\partial}{\partial x}\,
 \\
 \hat G_1 &=
    \frac{1}{4 \Omega^2}\, (-4\omega^2 +\gamma^2 \cos 2 \Omega t 
      + 2 \gamma \Omega \sin 2\Omega t) \,,
\\
 \hat G_2 &=
    -\frac{\gamma}{\Omega^2}\, \sin^2 \Omega t \,,
\end{align*}
so that they close the seven-dimensional algebra: 
\begin{align*}
  \left[\hat X, \hat P \right] &= i\hbar \hat{I}& 
  \left[\hat{\tilde Q}, \hat \Pi \right] &= 2 i \hbar \hat G_1 + i \hbar\hat{I}
\\
  \left[\hat X, \hat{\tilde Q} \right] &= \frac{i\hbar}{m} \hat G_2&
  \left[\hat X, \hat \Pi \right] &= i\hbar \hat G_1
\\
  \left[\hat{\tilde Q}, \hat P \right] &= -i\hbar \hat G_1 
                   + i\hbar \gamma \hat G_2&
  \left[\hat P, \hat \Pi \right] &= - i \hbar m \omega^2 \hat G_2
\\
  \left[\hat H, \hat X \right] &= \frac{i\hbar}{m} \hat \Pi & 
  \left[\hat H, \hat P \right] &= 2 i \hbar m \omega ^2 \hat X 
                     - i \hbar m \omega^2 \hat{\tilde Q}
\\
  \left[\hat H, \hat{\tilde Q} \right] &= -2 i\hbar \gamma \hat X 
      -\frac{i\hbar}{m} \hat P + i\hbar \gamma \hat{\tilde Q} &
  \left[\hat H, \hat \Pi \right] &= -3 i\hbar m \omega^2 \hat X 
     + 2 i\hbar m \omega^2 \hat{\tilde Q} - i\hbar \gamma \hat \Pi
\\
  \left[\hat H, \hat G_1 \right] &= -i\hbar \gamma \hat G_1 
                   + 2 i\hbar \omega^2 \hat G_2&
  \left[\hat H, \hat G_2 \right] &= - 2 i \hbar \hat G_1
                 + i \hbar \gamma \hat G_2 - 2 i \hbar \hat{I}
\end{align*}

We see that this algebra
corresponds to a centrally extended algebra. The central extensions determine
the actual basic conjugated pairs and classify possible quantizations.
The operators $\hat{\tilde Q}$ and $\hat{\Pi}$ (plus $\hat{I}$) expand a
Heisenberg-Weyl subalgebra, and $\hat{H}, \hat{G}_1$ and $\hat{G}_2$ expand a
2-D affine algebra (with $\hat{H}$ acting as dilations).
However, in this realization $\hat{\tilde Q}$ and $\hat{\Pi}$ are not basic
(this can be seen as an anomaly), and $\hat{H}$ and $\hat{G}_2$ are basic,
resulting in time being a canonical variable. Clearly, this is not satisfactory,
and an alternative description should be looked for.

Our strategy here is to consider other possible quantizations of the un-extended
algebra.
A detailed study of the (projective) representations of the enlarged
(7+1) dimensional Lie algebra (that is, the possible central extensions) is
going to show that there are three relevant kinds of representations,
describing systems with different degrees of freedom. 

Thinking of the algebra above as an abstract Lie algebra, it can be shown that
a parameter $k$ controls the central extensions which are allowed by the Jacobi
identity: 

\begin{align*}
  \left[\hat X, \hat P \right] &= i\hbar \hat{I}& 
  \left[\hat{\tilde Q}, \hat \Pi \right] &= 2 i \hbar \hat G_1 + i \hbar
k \hat{I} 
\\
  \left[\hat X, \hat{\tilde Q} \right] &= \frac{i\hbar}{m} \hat G_2&
  \left[\hat X, \hat \Pi \right] &= i\hbar \hat G_1
\\
  \left[\hat{\tilde Q}, \hat P \right] &= -i\hbar \hat G_1 
                   + i\hbar \gamma \hat G_2 + i\hbar (1-k)\hat{I}&
  \left[\hat P, \hat \Pi \right] &= - i \hbar m \omega^2 \hat G_2
\\
  \left[\hat H, \hat X \right] &= \frac{i\hbar}{m} \hat \Pi & 
  \left[\hat H, \hat P \right] &= 2 i \hbar m \omega ^2 \hat X 
                     - i \hbar m \omega^2 \hat{\tilde Q}
\\
  \left[\hat H, \hat{\tilde Q} \right] &= -2 i\hbar \gamma \hat X 
      -\frac{i\hbar}{m} \hat P + i\hbar \gamma \hat{\tilde Q}&
  \left[\hat H, \hat \Pi \right] &= -3 i\hbar m \omega^2 \hat X 
     + 2 i\hbar m \omega^2 \hat{\tilde Q} - i\hbar \gamma \hat \Pi
\\
  \left[\hat H, \hat G_1 \right] &= -i\hbar \gamma \hat G_1 
                   + 2 i\hbar \omega^2 \hat G_2&
  \left[\hat H, \hat G_2 \right] &= - 2 i \hbar \hat G_1
                 + i \hbar \gamma \hat G_2 - i\hbar (1+k) \hat{I}
\end{align*}

It is convenient to perform the shift: 
\[
  \hat Q \equiv - \hat{\tilde Q} + (1-k) \hat X\,,
\]
so that the actual degrees of freedom diagonalize:
\begin{align*}
  \left[\hat X, \hat P \right] &= i\hbar \hat{I}& 
  \left[\hat Q, \hat \Pi \right] &= - i \hbar (k+1) \hat G_1 - i \hbar k \hat{I}
\\
  \left[\hat X, \hat Q \right] &= -\frac{i\hbar}{m} \hat G_2&
  \left[\hat X, \hat \Pi \right] &= i\hbar \hat G_1
\\
  \left[\hat Q, \hat P \right] &= i\hbar \hat G_1 
                   - i\hbar \gamma \hat G_2 &
  \left[\hat P, \hat \Pi \right] &= - i \hbar m \omega^2 \hat G_2
\\
  \left[\hat H, \hat X \right] &= \frac{i\hbar}{m} \hat \Pi & 
  \left[\hat H, \hat P \right] &= i \hbar m \omega ^2 (1+k) \hat X 
                     + i \hbar m \omega^2 \hat Q
\\
  \left[\hat H, \hat Q \right] &=
      i\hbar \gamma (1+k) \hat X +\frac{i\hbar}{m} \hat P &
%        + i\hbar \gamma \hat q_k + \frac{i \hbar}{m}(1-k)\hat \pi
  \left[\hat H, \hat \Pi \right] &= - i\hbar m \omega^2 (2k+1)\hat X 
%       - 2 i\hbar m \omega^2 \hat q_k - i\hbar \gamma \hat \pi
\\
      & \qquad + i\hbar \gamma \hat Q + \frac{i \hbar}{m}(1-k)\hat \Pi  \nn &
      & \qquad - 2 i\hbar m \omega^2 \hat Q - i\hbar \gamma \hat \Pi    \nn
\\
  \left[\hat H, \hat G_1 \right] &= -i\hbar \gamma \hat G_1 
                   + 2 i\hbar \omega^2 \hat G_2  &
  \left[\hat H, \hat G_2 \right] &= - 2 i \hbar \hat G_1
                 + i \hbar \gamma \hat G_2 - i\hbar (1+k) \hat{I}
\end{align*}

We can see that the representations of this algebra include:
\begin{itemize}
\item  For arbitrary $k$, a generic family with 3 degrees of freedom:
$(\hat{X},\hat{P})$,
$(\hat{Q},\hat{\Pi})$ and
$(\hat{H},\hat{G}_2)$, then time being a canonical variable.

\item For $k=1$, already
described, an anomalous family with 2 degrees of freedom:
$(\hat{X},\hat{P})$
and $(\hat{H},\hat{G}_2)$, then time being a canonical variable.

\item For $k=-1$, a family with 2 degrees of freedom: $(\hat{X},\hat{P})$ and
$(\hat{Q},\hat{\Pi})$.

\end{itemize}

Clearly, the interesting case is the third one, since it contains two
degrees of freedom and time is not a canonical variable. Its algebra is given
by:
\begin{align*}
  \left[\hat{X}, \hat{P} \right] &= i\hbar\hat{I} & 
  \left[\hat{Q}, \hat{\Pi} \right] &=  i \hbar \hat{I}\nn
\\
  \left[\hat{X}, \hat{Q} \right] &= -\frac{i\hbar}{m} \hat{G}_2&
  \left[\hat{X},  \hat{\Pi} \right] &= i\hbar \hat{G}_1 \nn
\\
  \left[\hat{Q},  \hat{P} \right] &= i\hbar \hat{G}_1 
                   - i\hbar \gamma \hat{G}_2 &
  \left[\hat{P},  \hat{\Pi} \right] &= - i \hbar m \omega^2 \hat{G}_2 \nn
\\
  \left[\hat{H}, \hat{X} \right] &= \frac{i\hbar}{m} \hat{\Pi} & 
  \left[\hat{H}, \hat{P} \right] &= i \hbar m \omega^2  \hat{Q}  \nn
\\
  \left[\hat{H}, \hat{Q}\right] &= \frac{i\hbar}{m} (\hat{P}+2\hat{\Pi}) 
       &
  \left[\hat{H}, \hat{\Pi} \right] &= i\hbar m \omega^2 (\hat{X}-2 \hat{Q})  \nn
\\ & + i\hbar \gamma \hat{Q} & &
      - i\hbar \gamma  \hat{\Pi}  \nn
\\
  \left[\hat{H}, \hat{G}_1 \right] &= -i\hbar \gamma \hat{G}_1 
                   + 2 i\hbar \omega^2 \hat{G}_2&
  \left[\hat{H}, \hat{G}_2 \right] &= - 2 i \hbar \hat{G}_1
                 + i \hbar \gamma \hat{G}_2  \,.  
\end{align*}

In this case the operators $\hat{G}_1$ and $\hat{G}_2$ are gauge (they commute
with the basic couples $(\hat{X},\hat{P})$ and $(\hat{Q},\hat{\Pi})$) and
therefore are represented trivially.

 The effective dimension of the algebra is
$5+1$: $(\hat{X},\hat{P})$, $(\hat{Q},\hat{\Pi})$, $\hat{H}$ and $\hat{I}$.
\begin{align*}
  \left[\hat{X}, \hat{P} \right] &= i\hbar \hat{I} & 
  \left[\hat{Q}, \hat{\Pi} \right] &=  i \hbar\hat{I} \nn
\\
  \left[\hat{X}, \hat{Q} \right] &= 0&
  \left[\hat{X},  \hat{\Pi} \right] &= 0\nn
\\
  \left[\hat{Q},  \hat{P} \right] &= 0 &
  \left[\hat{P},  \hat{\Pi} \right] &= 0 \nn
\\
  \left[\hat{H}, \hat{X} \right] &= \frac{i\hbar}{m} \hat{\Pi} & 
  \left[\hat{H}, \hat{P} \right] &= i \hbar m \omega^2  \hat{Q}  \nn
\\
 \left[\hat{H}, \hat{Q}\right] &= \frac{i\hbar}{m} (\hat{P}+2\hat{\Pi}) 
       &
  \left[\hat{H}, \hat{\Pi} \right] &= i\hbar m \omega^2 (\hat{X}-2 \hat{Q})  \nn
\\ & \,\,\,\,\, + i\hbar \gamma \hat{Q} & &
     \,\,\,\,\, - i\hbar \gamma  \hat{\Pi} \,.
\end{align*}

Here $\hat{H}$ is not a basic operator, and can be
written in terms of the basic ones in an irreducible representation:
\[
  \hat{H} = -\frac{1}{m} \hat{\Pi} \hat{P}- \frac{\gamma}{2}( \hat{Q} \hat{\Pi}
+ \hat{\Pi} \hat{Q}) - \frac{\hat{\Pi}^2}{m}+ m
\omega^2 \hat{X} \hat{Q} 
              - m \omega^2 \hat{Q}^2\,.
\]

The classical version of the  Hamiltonian is:
\begin{equation}
  H = -\frac{1}{m} \Pi P - \gamma Q \Pi
 - \frac{\Pi^2}{m}+ m
\omega^2 X Q  - m \omega^2 Q^2\,.
\label{eq:Hourclasico}  
\end{equation}

% 
% 
% In fact, if we fix the value of the
% commutator xp, .... Only the value $k=-1$ makes the Hamiltonian non-dynamical,
% as physically expected, and reduces the number of degrees of freedom to
% 2+2\footnote{There are other possibilities that cause the effective number of
% degreesof freedom to be 2+2 instead of 3+3, due to the presence of anomalies.
% This requires further study  that will be done  in the  announced  forthcoming
% paper  \cite{forthcoming}.}
% As it can be checked in the algebra \ref{}, the value $k=-1$ implies the
% ``ghost'' nature of the new degree of freedom. 
% Diagonalization of the symplectic form. 
% This actually overcome the difficulties expressed in \cite{Dekker_report} 
% to reconcile the commutation relations with the structure of the Hamiltonian. 
% This is rooted in having adopted the algebra of symmetries as the basic object
% describing the physical system and recovering the rest of objects from it. In
% particular the Hamiltonian.

\subsection{Bateman's system}
%%%%%%%%%%%%%%%%%%%%%%%%%%%%%%%%%%%%%%%%%%%%%%%%%%%%%%%%%%%%%%%%%%%%%%%%%%%%%%%%
%%%%%%%5
%%%%%%%%%%%%%%%%%%%%%%%%%%%%%%%%%%%%%%%%%%%%%%%%%%%%%%%%%%%%%%%%%%%%%%%%%%%%%%%%
%%%%%%%%%

The classical Hamiltonian \eqref{eq:Hourclasico} can be transformed, using the
linear, constant, canonical transformation:
\begin{align}
X &= \frac{m  \omega ^2 y -( p_y+m  \frac{\gamma}{2}x ) i \Omega }{m
\omega \sqrt{- \gamma i \Omega }}\nn\\
P&= \frac{\omega  (  p_x -m \frac{\gamma}{2} y  +  m x i\Omega )}{
\sqrt{- \gamma i \Omega }}\nn \\
Q &= \frac{ m  \omega ^2 y -(  p_y - m \frac{\gamma}{2} x ) 
 i\Omega }{ m
\omega  \sqrt{- \gamma  i\Omega }}\nn \\
\Pi &= -\frac{\omega  ( p_x +m \frac{\gamma}{2} y  +  m x i\Omega )}{
\sqrt{- \gamma  i\Omega }}\,,
\end{align}
into the Bateman dual Hamiltonian
\be
H_B=\frac{ p_x   p_y}{m}+\frac{\gamma}{2}( y p_y  -x p_x)+m \Omega ^2 x y \,,
\ee
that describes a damped particle $(x,p_x)$ and its time reversal $(y,p_y)$:
\be
\ddot{x}+\gamma\dot{x}+\omega^{2}x=0\,,\qquad
\ddot{y}-\gamma\dot{y}+\omega^{2}y=0\,.
\ee

The quantum Bateman Hamiltonian is:
\be
\hat{H}_B=\frac{ \hat{p}_x   \hat{p}_y}{m}+\frac{\gamma}{2}( \hat{y} \hat{p}_y 
-\hat{x} \hat{p}_x)+m \Omega
^2 \hat{x}\hat{y} \,,
\ee
and the Schr\"odinger equation for the Bateman's system is given by\footnote{The
Bateman system admits an equivalent description in terms of a real,
first-order, Schr\"odinger equation (see
Subsection \ref{firstorderpola})}:
\be
i\hbar\frac{\partial\phi(x,y,t)}{\partial t}
=\left[-\frac{\hbar^2}{m}\frac{\partial^2}{\partial x\partial y}
-i\hbar \frac{\gamma}{2}(y\parcial{y} - x\parcial{x})+m \Omega ^2 x
y\right]\phi(x,y,t)\,.
\ee

The system is conservative, so our objective of including time
evolution among the symmetries has been accomplished, at the cost of including a
new degree of freedom. $\hat{H}_B$ closes a 5+1 dimensional
algebra with $(\hat{x},\hat{p}_x)$ and
$(\hat{y},\hat{p}_y)$:
\begin{align}
  \left[\hat{x}, \hat{p}_x \right] &= i\hbar \hat{I} & 
  \left[\hat{y}, \hat{p}_y \right] &=  i \hbar\hat{I} \nn
\\
  \left[\hat{x}, \hat{y} \right] &= 0&
  \left[\hat{x},  \hat{p}_y \right] &= 0 \nn
\\
  \left[\hat{y},  \hat{p}_x \right] &= 0 &
  \left[\hat{p}_x,  \hat{p}_y \right] &= 0 \nn
\\
  \left[\hat{H}_B, \hat{x} \right] &= \frac{i\hbar}{m}(- \hat{p}_y
+m \frac{\gamma}{2}\hat{x})& 
  \left[\hat{H}_B, \hat{p}_x \right] &= i \hbar( -\frac{\gamma}{2} \hat{p}_x+ m
\Omega^2 \hat{y} ) \nn
\\
 \left[\hat{H}_B, \hat{y}\right] &= \frac{i\hbar}{m}(- \hat{p}_x
-m \frac{\gamma}{2}\hat{y})
       &
  \left[\hat{H}_B, \hat{p}_y \right] &= i \hbar( \frac{\gamma}{2} \hat{p}_y+ m
\Omega^2 \hat{x} ) \,.
\label{batealgebra}
\end{align}

However, it has been argued that the quantum Bateman's system possesses
inconsistencies, like complex eigenvalues and non-normalizable eigenstates.
Chru\'sci\'nski \& Jurkowski \cite{Chruscinski} showed that
$\hat{H}_B$ has real,
continuous spectrum (we will provide a prove of this in
Subsection~\ref{firstorderpola}), and that the complex eigenvalues are
associated with
resonances, which in last instance are the responsible of dissipation.

\subsection{Bateman's group law}
\label{bategroup}

% \textbf{Hay que cambiar el signo de t y gamma!!!}

The Lie algebra \eqref{batealgebra} can be exponentiated to give a Lie group,
whose group law we have found to be:
%
% Pongo la ley y los campos en el coborde bueno para la pola de primer orden
% (pero también está todo para la de segundo y para el caso ``simétrico''):
\begin{align*}
  t'' &= t'+t 
 \\
  x'' &= x + x' \, e^{-\frac{\gamma t}{2}} \, \cos \Omega t 
          + \frac{p_y'}{m \Omega} \, e^{-\frac{\gamma t}{2}} \, \sin \Omega t 
  \\
  y'' &= y + y' \, e^{\frac{\gamma t}{2}} \, \cos \Omega t 
          + \frac{p_x'}{m \Omega} \, e^{\frac{\gamma t}{2}} \, \sin \Omega t
  \\
  p_x'' &= p_x + p_x'  \, e^{\frac{\gamma t}{2}} \, \cos \Omega t 
          - m \Omega \, y' \, e^{\frac{\gamma t}{2}} \, \sin \Omega t
  \\
  p_y'' &= p_y + p_y'  \, e^{-\frac{\gamma t}{2}} \, \cos \Omega t 
          - m \Omega \, x' \, e^{-\frac{\gamma t}{2}} \, \sin \Omega t
  \\
  \zeta'' &= \zeta' \zeta e^{\frac{i}{\hbar}
     \lbrace  y' p_y \, e^{\frac{\gamma t}{2}} \, \cos \Omega t 
     - x \, p_x' \, e^{\frac{\gamma t}{2}} \, \cos \Omega t
     + m \Omega \, x \, y' \, e^{\frac{\gamma t}{2}} \, \sin \Omega t
     + \frac{1}{m \Omega} \, p_x' p_y \, e^{\frac{\gamma t}{2}} \, \sin \Omega t
     \rbrace} \,.
\end{align*}

This group law had not been considered previously in the literature, up to the
author's knowledge. 

The corresponding left-invariant vector fields can be computed: 
\begin{align*}
   \begin{split} 
  \tilde X_t^L &=
       \frac{\partial}{\partial t} 
        + (-\frac{\gamma}{2} x +\frac{p_y}{m}) \frac{\partial}{\partial x}
        + (\frac{\gamma}{2} y +\frac{p_x}{m}) \frac{\partial}{\partial y}
       \\
        & \qquad \qquad \qquad
        + (\frac{\gamma}{2} p_x - m \Omega^2 y) \frac{\partial}{\partial p_x}
        + (-\frac{\gamma}{2} p_y - m \Omega^2 x) \frac{\partial}{\partial p_y}
   \end{split}
\\
  \tilde X_x^L &= \frac{\partial}{\partial x} - \frac{p_x}{\hbar} \, \Xi
\\
  \tilde X_y^L &= \frac{\partial}{\partial y}
\\
  \tilde X_{p_x}^L &= \frac{\partial}{\partial p_x}
\\
  \tilde X_{p_y}^L &= \frac{\partial}{\partial x} + \frac{y}{\hbar} \, \Xi \,, 
\end{align*}
and also the right-invariant ones: 
\begin{align*}
  \tilde X_t^R &=  \frac{\partial}{\partial t} 
\\
  \tilde X_x^R &= 
    e^{-\frac{\gamma t}{2}} \, \cos \Omega t \frac{\partial}{\partial x}
   - m \Omega \, e^{-\frac{\gamma t}{2}} \, 
                        \sin \Omega t \frac{\partial}{\partial p_y}
\\
  \tilde X_y^R &= 
    e^{\frac{\gamma t}{2}} \, \cos \Omega t \frac{\partial}{\partial y}
   - m \Omega \, e^{\frac{\gamma t}{2}} \, 
                        \sin \Omega t \frac{\partial}{\partial p_x}
   + \frac{1}{\hbar} 
    (p_y \, e^{\frac{\gamma t}{2}} \, \cos \Omega t
   + m \Omega \, x \, e^{\frac{\gamma t}{2}} \, \sin \Omega t) \, \Xi
\\
  \tilde X_{p_x}^R &= 
    e^{\frac{\gamma t}{2}} \, \cos \Omega t \frac{\partial}{\partial p_x}
   + \frac{1}{m \Omega} \, e^{\frac{\gamma t}{2}} \, 
                        \sin \Omega t \frac{\partial}{\partial y}
   - \frac{1}{\hbar} 
    (x \, e^{\frac{\gamma t}{2}} \, \cos \Omega t
  - \frac{1}{m \Omega} \, p_y \, e^{\frac{\gamma t}{2}} \, \sin \Omega t) \, \Xi
\\
  \tilde X_{p_y}^R &= 
    e^{-\frac{\gamma t}{2}} \, \cos \Omega t \frac{\partial}{\partial p_y}
   + \frac{1}{m \Omega} \, e^{-\frac{\gamma t}{2}} \, 
                        \sin \Omega t \frac{\partial}{\partial x}\,.
\end{align*}

These vector fields close the Lie algebra \eqref{batealgebra}, provided
that obvious identifications are made.

\section{A note on the quantization of the Bateman's dual system}
\label{sec:arnold.quantbat}

\subsection{First-order Schr\"odinger equation}
\label{firstorderpola}

Usual Canonical Quantization leads to either position space or momentum space
representation and a corresponding second-order Sch\"odinger equation. However,
inspecting the Bateman's Lie algebra, it is possible to check that a full
first-order polarization exists: 
\[
  \mathcal P = \langle \tilde X_y^L,\tilde X_{p_x}^L ,\tilde X_t^L \rangle\,.
\]

The first two polarization conditions determine that wave functions are
($U(1)$-functions) depending only on $(x, p_y, t)$. The last polarization
equation $ \tilde X_t^L \psi = 0$ determines the condition on functions on the
reduced space, $\phi(x, p_y, t)$: 
\[
  \frac{\partial \phi}{\partial t} =  
        - (-\frac{\gamma}{2} x +\frac{p_y}{m}) \frac{\partial \phi}{\partial x}
%        \\
%         & \qquad \qquad \qquad
    - (-\frac{\gamma}{2} p_y - m \Omega^2 x) \frac{\partial \phi}{\partial p_y}
\,.
\]

We have arrived at a first-order partial differential equation that must be
interpreted as a first-order Schr\"odinger equation in a mixed representation
position-momentum. In fact, the same result can be obtained performing
Canonical Quantization for the Bateman Hamiltonian $\hat H_B$ in this mixed
representation. Let us emphasize that this has been strongly suggested by the
group structure and the GAQ algorithm. 

The corresponding time-independent Schr\"odinger equation is written: 
\[
   (\frac{\gamma}{2} x -\frac{p_y}{m}) \frac{\partial \phi}{\partial x}
%        \\
%         & \qquad \qquad \qquad
    + (\frac{\gamma}{2} p_y + m \Omega^2 x) \frac{\partial \phi}{\partial p_y}
= E \phi
\,.
\]
The general solution of this equation can be found in terms of the complex
variable $z\equiv p_y + i m \Omega x$:
\[
  \phi(z) = \Bigl(\frac{z}{z^*}\Bigr)^{\frac{E}{2 \hbar \Omega}} 
            f\bigl( z z^* \Bigl( \frac{z}{z^*}\Bigr)^{\frac{i \gamma}{2 \Omega}}
\bigr)\,,
\]
where $f$ is an arbitrary function of its argument. 

Let us focus in the case of underdamping, where $\Omega$ is real. We must
determine whether $\phi$ is well defined. To this end, we assume that
$f$ is a power of its argument, $ (z z^*)^{\tilde \lambda} \Bigl(
\frac{z}{z^*}\Bigr)^{\frac{i \gamma \tilde \lambda}{2 \Omega}}$, 
%so that we are left with another arbitrary function $g$ of the argument. 
and we write: 
\[
  \phi(z) = \Bigl(\frac{z}{z^*}\Bigr)^{\frac{E}{2 \hbar \Omega}} 
            (z z^*)^{\tilde \lambda} \Bigl(
           \frac{z}{z^*}\Bigr)^{\frac{i \gamma \tilde \lambda}{2 \Omega}}
%             g\bigl( z z^* \Bigl( \frac{z}{z^*}\Bigr)^{\frac{i \gamma}{2
% \Omega}}\bigr)
  = \Bigl(\frac{z}{z^*}\Bigr)^{\frac{E+ i \hbar \tilde \lambda \gamma}{2 \hbar
\Omega}} (z z^*)^{\tilde \lambda}
% g\bigl( z z^* \Bigl( \frac{z}{z^*}\Bigr)^{\frac{i \gamma}{2 \Omega}}\bigr)
\,.
\]
Now, 
% the function $g$ and the rest of 
the wave function have to be well-defined. 
%This restricts the possible $g$'s. But it also 
This imposes a ``quantization'' condition on the spectrum. 

On the one hand, recall that $z z^*$ is real. For $\phi$ to be at least
Dirac-delta normalizable, $\tilde\lambda$ must be chosen to be pure imaginary: 
\[
  \tilde \lambda = i \lambda\,, \qquad \lambda \in \mathbb R\,.
\]
On the other hand, $\frac{z}{z^*}$ is a pure phase, with twice the argument of
$z$. The exponent of $\frac{z}{z^*}$ must be half-integer so that we can get
a well-defined function of $z$: 
\[
  E - \hbar \gamma \lambda = n \hbar \Omega \quad \Rightarrow\quad 
\boxed{E = n \hbar \Omega + \lambda \hbar \gamma}\,.
\]
That is, we obtain a spectrum which has an integer part and a continuous part. 

These results coincide with those in \cite{Chruscinski}, although here they are
obtained in a quicker and neater way. The reason is that they quantize angular
variables and hence the basic operator ``multiply by the angle'' is not
defined. We have avoided this problem. However, it is somewhat surprising that
both spectrums coincide.

The fact that the spectrum of $\hat H_B$ has an integer part and a continuous
part suggests that $\hat H_B$ can be split into a compact operator (of the
harmonic oscillator type) and another operator with an unbounded, continuous
spectrum. This splitting should be found in the Lie algebra of the Bateman's
group, and is under investigation. 

\subsection{Back to Caldirola-Kanai system}
\label{back2caldi}

Historically, Bateman firstly derived  $H_B$, and later Caldirola
and Kanai obtained $H_{DHO}$ using time-dependent canonical transformations.
Here we have gone the opposite way, started from $H_{DHO}$ and derived $H_B$
closing a finite Lie algebra.
Now we wonder if we can do the way back  to the Caldirola-Kanai system. The
answer, again, is positive, and can be achieved by using constraints.
To know how to proceed, let us analyse first the classical case.

Classically, Bateman's system  and a pair of dual Caldirola-Kanai systems share
the
same second-order equations of motion. If we impose them to share the
first-order,
Hamilton equations, the following constraint must be satisfied:
\begin{align}
y&=\frac{\omega^2}{\Omega^2} e^{\gamma t}x+\frac{\gamma}{2m\Omega^2}p_x \nn \\
p_y&=e^{\gamma t}p_x+m\frac{\gamma}{2}x \,.
\end{align}

These constraints, although time dependent, preserve the equations of motion
since they are equivalent to a relation among initials constants:
\begin{align}
y_0&=\frac{\omega^2}{\Omega^2}x_0+\frac{\gamma}{2m\Omega^2}p_{x0}\nn \\
p_{y0}&=p_{x0}+m\frac{\gamma}{2}x_0\,.
\end{align}

%Classically, the Caldirola-Kanai system can be derived from the Bateman's
% system
%using these constraints.
These constraints can be seen to be of second-order type, besides being
time-dependent, therefore care should
be taken when imposing them: Dirac theory for constraints can be used or we can
embed the constraints
in a time-dependent canonical transformation before applying them.

But we are interested in the quantum derivation. Therefore we try to impose the
operator
constraints:
\begin{align}
\hat{y}-\frac{\omega^2}{\Omega^2}\hat{x}-\frac{\gamma}{2m\Omega^2}\hat{p}_x &=
0
\nn \\
\hat{p}_y-\hat{p}_x-m\frac{\gamma}{2}\hat{x} &= 0	\,,
\end{align}

\noindent but only one of them can be imposed, since the  operators at the lhs
of the equations canonically commute: they are of second
order type. At the quantum level, only one of them can be imposed, therefore we
must select one of them. If we impose the constraint,
\be
\hat{y}=\frac{\omega^2}{\Omega^2}\hat{x}+\frac{\gamma}{2m\Omega^2}\hat{p}_x\,,
\ee

\noindent the Hilbert space reduces to those functions verifying:
\be
\phi(x,y,t)=e^{\frac{i e^{- \gamma t } m  \Omega y  \text{Csc}^2(\Omega t)
\left( \gamma \Omega y \text{Cos}(2 \Omega t)+2 \left(\omega^2 e^{\gamma t } x'
- \Omega^2 y\right) \text{Sin}(2 \Omega t)\right)}{4\hbar \omega ^2 }}
\psi(x',t)\,,
\ee

\noi where  $x'=x+\frac{\Omega^2}{2 \omega^2} y e^{- \gamma t } \mu(t) $, and
$\mu(t)=(2 
-\frac{\gamma}{\Omega} \text{Cot}( \Omega t ))$. The Schr\"odinger equation for
the Bateman's system reduces to:
\be
i \hbar \dparcial{\psi(x',t)}{t} =\left[-\frac{\Omega^2  \hbar ^2}{2 m \omega
^2} e^{- \gamma t } \mu(t)
\frac{\partial^2\,}{\partial x'^2} 
 - \frac{1}{2} i\hbar x'  \Omega \mu(t) \parcial{x'}
+i \hbar\frac{\Omega^2}{\gamma}  (\mu(t)-2) \right]\psi(x',t) \,.
\ee

When constraints are imposed, not all the operators acting on the original
Hilbert space preserve the constrained
Hilbert space. The notion of ``good'' (usually denoted gauge-independent in
constrained gauge theories) operators as
those preserving the constrained Hilbert space naturally emerges.

In most of the cases ``good'' operators are characterized as those commuting
with the constraints 
(see \cite{halleff} for a detailed account of quantum constraints in a 
group-theoretical setting and a more general characterization of
``good'' operators). In this case they are:
\be
 \hat{p}_x+\frac{ 2m \omega ^2}{ \gamma } \hat{x}\qquad
\hat{p}_y-\frac{2 m \Omega ^2}{\gamma } \hat{x}\,.\label{operatasbuenos}
\ee

Note that  $\hat{H}_B$ (nor $i \hbar \parcial{t}$) is not among the ``good''
operators since it does not preserve the
constrained Hilbert space. Therefore, time invariance is lost in the process of
going from the Bateman's system to the
Caldirola-Kanai system due to the very nature of the constraints imposed.

Now let us perform the transformation
\be
\psi(x',t)=e^{-i \frac{m \omega ^2}{\hbar\Omega} x'^2
f(t)}g(t)\chi(\kappa,\tau)\,,
\ee

\noi where
\be
f(t)=-\frac{e^{\gamma t} }{4 \Omega\mu(t)^2 \tau'(t)}
 \left( (-\gamma (2+\text{Cos}(2  \Omega t )) 
 +2  \Omega \text{Sin}(2 
\Omega t)) \tau'(t) 
- \gamma  \mu(t)  \tau'(t)^2 
+
 \mu(t)\tau''(t)\right)
\ee
\be
g(t)=e^{-\frac{1}{4} \gamma 
\tau}\left(-\frac{\tau'(t)}{\Omega \text{Sin}^2(\Omega t)\mu(t)}\right)^{1/4} 
\ee
\be
\kappa=x' e^{\frac{\gamma}{2} (t
-\tau)}\frac{\omega}{\Omega}\sqrt{\frac{ \tau'(t)}{\mu(t)}}
\ee
\be
\tau(t)=\frac{1}{\Omega }\text{ArcTan}\left[\frac{A
\frac{\gamma^2}{\Omega^2}}{\mu(t)^2}\right]\,,\,A\in \mathbb{R}-\{0\}\,.
\ee

The Schr\"odinger equation finally  transforms into:
\be
i\hbar\frac{\partial}{\partial \tau}\chi(\kappa,\tau)=\left[
-\frac{\hbar^{2}}{2m} e^{-\gamma \tau}\frac{\partial^{2}}{\partial \kappa^{2}}
+\frac{1}{2}m\omega^{2}\kappa^{2} e^{\gamma \tau}\right]\chi(\kappa,\tau) \,,
\ee

\noi which is the Caldirola-Kanai equation in the variables $(\kappa,\tau)$.
Even more, the two independent operators (\ref{operatasbuenos}) preserving the
constrained Hilbert  space turn, 
under the previous transformation, to
the basic operators for the Caldirola-Kanai system $\hat{x}(t)$ and
$\hat{p}(t)$. Therefore, we have recovered
completely the Caldirola-Kanai system from the Bateman's system by imposing one
constraint.

It should be stressed that $\tau'(0)=0$, therefore the time transformation is
singular at the origin and there are two disconnected regions, one with $t>0$
and other with $t<0$. 
It also turns out that $sign(\tau)=sign(A)$, therefore choosing  appropriately
the sign of $A$ in each case we can map $t>0$ to $\tau>0$ and $t<0$ to 
$\tau<0$, respectively.

This kind of behavior coincides with the results of other authors (see
\cite{Chruscinski}) where, starting with 
the Bateman's system, they
obtain two subspaces ${\cal S}^{\pm}$ for which the restriction of the one
parameter group of unitary 
time-evolution operators $\hat{U}(t)=e^{-\frac{i}{\hbar}t\hat{H}_B}$ produces
two
semigroups of operators, for $t<0$ and $t>0$.

Therefore, starting from the quantum, conservative, Bateman's system we have
arrived to the quantum, time-dependent, Caldirola-Kanai system. All the
process we have performed can be schematically showed as:

\begin{equation}
\begin{array}{ccc}
 & \stackrel{\text{Constraint}}{\Rightarrow}& \\
\text{Bateman} &   & \text{Caldirola-Kanai}\\
t\in \mathbb{R} & &  t\in\mathbb{R}^+ \,\,{\rm or}\,\, t\in\mathbb{R}^- \\
\text{Conservative} & & \text{Dissipative} \\ 
 & \stackrel{\text{Closing algebra}}{\Leftarrow} &
\end{array}
\end{equation}

%\appendix
\section*{Appendix: Infinite-dimensional symmetry in the damped particle}
\label{infinitegroup}

In this appendix, we turn our attention to the damped particle as the
simplest case of physical system subjected to a dissipative force and perform a
similar analysis to that carried out in Subsection \ref{timesymmetry} for the
damped harmonic oscillator, forcing the introduction of the time symmetry. 

The basic operators for the damped particle (obtained as the $\omega\rightarrow 0$ of the damped harmonic oscillator, 
see eq. (\ref{eq:P_DHO}-\ref{eq:X_DHO}))
\begin{equation}
 \hat P = -i \hbar \frac{\partial}{\partial x},\quad 
 \hat X = x+\frac{i \hbar}{m \gamma}(1-e^{-\gamma t})\frac{\partial}{\partial x}
 \,.
 \label{eq:operatas DP} 
\end{equation}

Reminding the operators \eqref{eq:operatas DP}, and renaming $\hat P\equiv\hat
P_0$, we introduce operators $\hat P_n$ and $\hat Y_n$ ($n$ an integer)
\begin{align}
 \hat H_G &= i \hbar e^{\gamma t}\frac{\partial}{\partial t} & \hat H_{DP} &= i
\hbar \frac{\partial}{\partial t}\\
 \hat P_n &= -i \hbar e^{- \gamma n t}\frac{\partial}{\partial x} & \hat Y_n
&=
i e^{- \gamma n t}\\
\hat X &= x + \frac{i \hbar}{m \gamma}(1-e^{\gamma t})\frac{\partial}{\partial
x}\,.
\end{align}

It is interesting that they close an infinite-dimensional Lie algebra 
\begin{align*}
[\hat H_G,\hat P_n]& =-i \hbar \gamma n \hat P_{n-1}&
[\hat H_{DP},\hat P_n]& =-i \hbar \gamma n \hat P_{n}\\
[\hat H_G,\hat X    ]& =-i \frac{\hbar}{m}\hat P_0&
[\hat H_{DP},\hat X    ]& =-i \frac{\hbar}{m}\hat P_1\\
[\hat H_G,\hat H_{DP}]& =-i \hbar \gamma \hat H_G&
[\hat H_{DP},\hat Y_n]& =-i \hbar \gamma n \hat Y_{n}\\
[\hat H_G,\hat Y_n]& =-i \hbar \gamma n \hat Y_{n-1}&
[\hat X,\hat P_n]& = \hbar \hat Y_n
\end{align*}
(other commutators vanish) which has the (centrally extended) Galilei algebra as
a subalgebra, noting that $\hat Y_0 = i$ is the central generator. 

The generators in the right column, with $n=0,1$, also close a finite
dimensional
subalgebra in which $\hat H_{DP}$ is dynamical, conjugate of a combination of
$\hat P_1$ and $\hat Y_1$, together with the couple $\hat X$, $\hat P_0$. A
similar analysis to that of the damped harmonic oscillator is then possible,
recovering the corresponding results when $\omega=0$.

However, we have found that it is possible to enlarge the algebra to an
infinite-dimensional one, at least in the case of the damped particle, and 
new degrees of freedom arise. A deeper analysis of this matter is under
study.

\section*{Acknowledgments}

Work partially supported by the Fundaci\'on S\'eneca, Spanish MICINN and
Junta de Andaluc\'\i a under projects 08814/PI/08, FIS2008-06078-C03-01 
and FQM219-FQM1951, respectively. 

The authors wish to thank M. Calixto for useful discussions and comments.


\begin{thebibliography}{99}

\bibitem{Scully-Zubairy} Scully M.O., Zubairy M.S., \textit{Quantum Optics}
(Cambridge University Press, 1997). %\\[-20pt]

\bibitem{Breuer} Breuer H-P., Petruccione F., \textit{The Theory of Open
Quantum System} (Oxford University Press, 2002). %\\[-20pt]

\bibitem{Dittrich} Dittrich T. et al., \textit{Quantum Transport and
Dissipation} (Wiley-VCH, 1998). %\\[-20pt]

\bibitem{Caldirola} Caldirola P., Nuovo Cimento (1941) 18, 393 %\\[-20pt]

\bibitem{Kanai} Kanai E. (1948) Prog. Theor. Phys. 3, 440 %\\[-20pt]

\bibitem{Brittin} Brittin W.E., Phys. Rev. \textbf{77}, 396 (1950) %\\[-20pt]

\bibitem{Razavy_libro} Razavy M., \textit{Classical and Quantum Dissipative
Systems} (Imperial College Press, 2005). %\\[-20pt]

\bibitem{Dekker_articulo} Dekker H., Phys. Rev. A \textbf{16}, 2126 (1977)
%\\[-20pt]

\bibitem{Razavy} Razavy M., Z. Physik B \textbf{26}, 201 (1977) %\\[-20pt]

\bibitem{Schuch} Schuch D., Phys. Rev. A \textbf{55}, 935 (1997) 

\bibitem{Manko} Dodonov V.V., Man'ko V.I., Phys. Rev. A \textbf{20}, 550 (1979)
%\\[-20pt]

\bibitem{Dekker_report} Dekker H., Phys. Rep. \textbf{80}, 1 (1981) %\\[-20pt]

\bibitem{Martini} Martini R., Kersten P.H.M., J. Phys. A \textbf{16}, 455
(1983) %\\[-20pt]

\bibitem{Cervero-C} Cerver\'o J.M., Villarroel J., J. Phys. A \textbf{17}, 1777 
(1984) %\\[-20pt]

\bibitem{Cervero-Q} Cerver\'o J.M., Villarroel J., J. Phys. A \textbf{17}, 2963
(1984) %\\[-20pt]

\bibitem{schrogroup} U. Niederer, Helv. Phys. Acta, \textbf{45}, 802 (1972);\textbf{46}, 191 (1973);\textbf{47}, 167 (1974)

\bibitem{Magnus} Blanes S., Casas F., Oteo J.A., Ros J., Phys. Rep.
\textbf{470}, 151 (2009) %\\[-20pt]


\bibitem{Yeon1987} C-I. Um, K-H. Yeon and W.H. Kahng, J. Phys. A \textbf{20}, 611 (1987)


\bibitem{Um-Yeon} Um C-I., Yeon K-H., George T.F., Phys. Rep. \textbf{362},
63 (2002) %\\[-20pt]

\bibitem{Hone} Hone A.N.W., Senthilvelan M., J. Math. Phys. \textbf{50}, 102902
(2009) %\\[-20pt]

\bibitem{Streklas} Streklas A., Physica A, \textbf{385}, 124 (2007) %\\[-20pt]

\bibitem{arnoldnuestro} V. Aldaya, F. Coss\'{\i}o, J. Guerrero and F.F. L\'opez-Ruiz, J. Phys. A, \textbf{44}, 065302
(2011). arXiv:1010.5521


\bibitem{Arnold} Arnold V. I., \textit{Geometrical methods in the theoy of
ordinary differential equations}, (Springer, 1998). %\\[-20pt]

\bibitem{Lutzky} Lutzky M. J. Phys. A, \textbf{11}, 249 (1978) %\\[-20pt]

\bibitem{Takagi} Takagi S., Prog. Theor. Phys. \textbf{84}, 1019 (1990)
%\\[-20pt]

\bibitem{Huang-Wu} Huang M-C., Wu M-C., Chin. J. Phys. \textbf{36}, 566 (1998)
%\\[-20pt]

\bibitem{Kanasugi} Kanasugi H., Okada H., Prog. Theor. Phys. \textbf{93}, 949
(1995) %\\[-20pt]

\bibitem{Bateman} Bateman H., Phys. Rev. \textbf{38}, 815 (1931) %\\[-20pt]

\bibitem{Tikochinsky} Y. Tikochinsky, J.Math. Phys. \textbf{19}, 888 (1978)

\bibitem{Fesh-Tiko} H. Feshbach and Y. Tikochinsky, in \textit{A Festschrift for I.I. Rabi}, vol. \textbf{38}, p. 44,
Trans. New York Ac. Sc. Ser. 2. (1977).


\bibitem{Feschbach_libro} P.M. Morse and H. Feshbach, \textit{Methods of Theoretical Physics}, vol. I, p. 298,
McGraw-Hill, New York (1953)

\bibitem{Celeghini} Celeghini E., Rasetti M., Vitielo G., Ann. Phys.
\textbf{215}, 156 (1992) %\\[-20pt]

\bibitem{Chruscinski} Chru\'sci\'nski D., Jurkowski J., Ann. Phys.
\textbf{321}, 854 (2006) %\\[-20pt]

\bibitem{Blasone} Blasone M., Jizba P., Ann. Phys. \textbf{312}, 312 (2004)
%\\[-20pt]

\bibitem{discretebasis} J. Guerrero, F.F. L\'pez-Ruiz, V. Aldaya and F. Coss\'{\i}o,\textit{Discrete basis of localized
quantum states for the free particle}. arXiv:1010.5525



%\bibitem{forthcoming} Forthcoming paper %\\[-20pt]

\bibitem{key-3} Aldaya V., de Azc\'arraga J.A., J. Math. Phys.
\textbf{23}, 1297  (1982) %\\[-20pt]

\bibitem{key-4} Aldaya V., Guerrero J. Rep. Math. Phys. \textbf{47}, 213 (2001)
%\\[-20pt]

\bibitem{key-5} Aldaya V., Calixto M., Guerrero J., L\'opez-Ruiz F. F.,  J.
Nonl. Math. Phys. \textbf{15}, 1 (2008) %\\[-20pt]

\bibitem{key-6} Aldaya V., Navarro-Salas J., Bisquert J., Loll R.,  J.Math.
Phys. \textbf{33}, 3087 (1992) %\\[-20pt]

\bibitem{halleff} V. Aldaya, M. Calixto and J. Guerrero, Commun.Math. Phys., \textbf{178}, 399 (1996)

\end{thebibliography}
\end{document}